\documentclass[preprint2]{aastex}

\newcommand{\lae}{\mathrel{<\kern-1.0em\lower0.9ex\hbox{$\sim$}}}
\newcommand{\gae}{\mathrel{>\kern-1.0em\lower0.9ex\hbox{$\sim$}}}
\shorttitle{Hard X-ray Spectrum of NGC 2992 and NGC 3081}
\shortauthors{Beckmann, Gehrels \& Tueller}

\begin{document}
   \title{The 1 keV to 200 keV X-ray Spectrum of NGC~2992 and NGC~3081}

%   \subtitle{Results from INTEGRAL, Swift, and BeppoSAX}

   \author{V. Beckmann\altaffilmark{1},}
      \affil{INTEGRAL Science Data Centre, Ch. d'\`Ecogia 16, 1290
      Versoix, Switzerland}
       \email{Volker.Beckmann@obs.unige.ch}

\author{N. Gehrels, \& J. Tueller}
      \affil{Astrophysics Science Division, NASA Goddard Space Flight Center, Code 661, Greenbelt, MD 20771, USA}
    \altaffiltext{1}{also with the Department of Physics, University of Maryland Baltimore County, 1000 Hilltop Circle, Baltimore, MD 21250, USA}
%   \altaffiltext{2}{also with the INTEGRAL Science Data Centre, Ch. d'\'Ecogia 16, 1290 Versoix, Switzerland}
% \and
% INTEGRAL Science Data Centre, Ch. d'\'Ecogia 16, 1290 Versoix, Switzerland\\
%             \email{c.ptolemy@hipparch.uheaven.space}
%             \thanks{The university of heaven temporarily does not
%                     accept e-mails}

\begin{abstract}
   The Seyfert 2 galaxies NGC~2992 and NGC~3081 have been observed by {\it INTEGRAL} and {\it Swift}. We report about the results and the comparison of the spectrum above 10 keV based on {\it INTEGRAL} IBIS/ISGRI, {\it Swift}/BAT, and {\it BeppoSAX}/PDS.
A spectrum can be extracted in the X-ray energy band ranging from 1 keV up to 200 keV.
   Although NGC~2992 shows a complex spectrum below 10 keV, the hard tail observed by various missions exhibits a slope of $\Gamma = 2$, independent on the flux level during the observation. No cut-off is detectable up to the detection limit around 200 keV. In addition, NGC~3081 is detected in the {\it INTEGRAL} and {\it Swift} observation and also shows an unbroken $\Gamma = 1.8$ spectrum up to 150 keV.
These two Seyfert galaxies give further evidence that a
high-energy cut-off in the hard X-ray spectra 
% is only observed in
%   a few exceptional cases and for most objects the cut-off is
is often located at energies $E_C \gg 100 \rm \, keV$. 
In NGC~2992 a constant spectral shape is observed over a hard X-ray luminosity variation by a factor of 11. This might indicate that the physical conditions 
%temperature 
of the emitting hot plasma are constant, while the amount of plasma 
%visible to the observer 
varies, due to long-term flaring activity.
\end{abstract}

   \keywords{galaxies: Seyfert  -- galaxies: active -- X-rays: galaxies
                -- galaxies: individual (NGC~2992, NGC~3081)
               }

%________________________________________________________________

\section{Introduction}

Active Galactic Nuclei (AGN) are the most prominent persistent X-ray sources in the extragalactic sky. 
The high energy spectrum of AGN in general has been controversial since AGN became observable in the X-ray band in the late 1960s. 
Early on, it was noticed that AGN exhibit a power law like spectrum in the X-rays. While the soft X-ray part is altered by absorption in the line of sight, the spectrum above 20~keV is hardly affected by absorption. The photon index in this energy range is usually of the order of $\Gamma \simeq 2$, but can range from $\Gamma \simeq 1$ to $\Gamma \simeq 3$ \cite{intagn} and appears in most cases to connect smoothly with
%be the unabsorbed continuation of 
the absorbed soft X-ray spectrum. 

%...

\object{NGC 2992} is one of the brightest AGN in the X-ray sky. This
nearby Seyfert 2 galaxy ({\it z} = 0.0077) has been observed by every
major X-ray satellite mission. It was first detected by the {\it Ariel
  5} Sky Survey Instrument as 2A~0943-140, an X-ray source which was
later identified to be NGC~2992 with an X-ray luminosity of $L_X = 2
\times 10^{43} \rm \, erg \, s^{-1}$ \cite{ariel}. A similar
luminosity was reported by the {\it High Energy Astrophysics
  Observatory} ({\it HEAO}) with $L_X = (1.4 \pm 0.4) \times 10^{43}
\rm \, erg \, s^{-1}$ \cite{griffiths}, by several observations with
{\it Einstein}, revealing $L_X = (0.6 - 1.4) \times 10^{43} \rm \, erg
\, s^{-1}$ \cite{Maccacaro}, and the {\it European X-ray Observatory
  Satellite} ({\it EXOSAT}) with $L_X = (0.3 - 1.2) \times 10^{43} \rm
\, erg \, s^{-1}$, $\Gamma = 1.6 {+0.6 \atop -0.4}$, and $N_H \simeq 5 \times
10^{21} \rm \, cm^{-2}$ ($N_{H,gal} = 8 \times 10^{20} \rm \,
cm^{-2}$) \cite{NGC2992EXOSAT}. 
The spectral properties at softer X-rays were slightly different, as shown by the next generation of X-ray
telescopes with higher spectral and angular resolution. The {\it
  Advanced Satellite for Cosmology and Astrophysics} ({\it ASCA})
measured in the $0.5 - 8.5 \rm \, keV$ energy band a photon index of  $\Gamma = 1.21 \pm 0.10$ and absorption of $N_H = 1.7 {+0.7 \atop -0.6} \times 10^{21} \rm \, cm^{-2}$ \cite{weaver}. They also detected a delayed response of the reflection component and derived a distance of $\sim 3.2 \rm \, pc$ for the reflecting material, which they assumed to be dense, neutral gas in the central region with column densities $N_H \simeq 10^{23} - 10^{25} \rm \, cm^{-2}$. 
The hard X-ray tail of NGC~2992 was first measured by {\it BeppoSAX} and revealed a photon index of $\Gamma \simeq 1.7$ up to an energy of $\sim 150 \rm \, keV$ with absorption by $N_H \simeq 10^{22} \rm \, cm^{-2}$ and an iron K$\alpha$ line at $E_{K\alpha} = (6.62 \pm 0.07) \rm \, keV$ \cite{gilli}. The line showed variable equivalent width (700~eV in 1997 and 147~eV in 1998), which was caused by variable continuum flux rather than by the iron line itself varying. 
Observations by {\it Chandra} allowed for the first time the disentangling of the complex spatial structure of NGC~2992. Colbert et al. (2005) showed that the X-ray emission can be described by three components: An AGN core with a photon index of $\Gamma = 1.86$, a cold Compton reflector with column density of $N_H \simeq 10^{22} \rm \, cm^{-2}$, and in the soft X-rays by a thermal plasma with $kT = 0.5 \rm \, keV$ and low abundance ($Z < 0.03 Z_\odot$). They also showed that the spectrum in the $2 - 10 \rm \, keV$ band can be described in first order by a simple power law with photon index $\Gamma = 0.91 \pm 0.02$ and absorption $N_H = (1.9 \pm 0.4) \times 10^{21} \rm \, cm^{-2}$, in which the flat power law accounts for the sum of the three components. Three observations by {\it Suzaku} at the end of 2005 in the $0.5 - 10 \rm \, keV$ band were modeled by Yaqoob et al. (2007) with a primary continuum ($\Gamma = 1.57 {+0.06 \atop -0.03}$) obscured by an absorber with $N_H = 8.0 {+0.6 \atop -0.5} \times 10^{21} \rm \, cm^{-2}$, plus an optically-thin thermal emission component with $kT = 0.66 {+0.09 \atop -0.06} \rm \, keV$, and an Fe K emission complex with a broad Fe K$\alpha$ component having $EW = 118 {+32 \atop -61} \rm \, eV$, and a narrow component with $EW = 163 {+47 \atop -26} \rm \, eV$. The source showed a 2~--~10~keV luminosity  of $L_X = 3 \times 10^{42} \rm \, erg \, s^{-1}$.  

Less X-ray observations have been performed on the Seyfert 2 galaxy NGC~3081 ($z = 0.00798$). This source has been first detected in X-rays  by the {\it Einstein} imaging instruments with a flux of $f_{(0.2 - 4.0 \rm keV)} = 8 \times 10^{-13} \rm \, erg \, cm^{-2} \, s^{-1}$ \cite{NGC3081EINSTEIN}. In the {\it ROSAT} All-Sky Survey \cite{RASS} the source was detected with a flux of $f_{(0.1 - 2.4 \rm keV)} = (2.4 \pm 0.9) \times 10^{-13} \rm \, erg \, cm^{-2} \, s^{-1}$. 
%A 35 ks observation of {\it ASCA} gave a flux of $f_{(2 - 10 \rm \, keV)} = 3.1 \times 10^{-12} \rm \, erg \, cm^{-2} \, s^{-1}$ and did not show any significant absorption ($N_{H,gal} = 4.6 \times 10^{20}$). 
{\it BeppoSAX} observed  NGC~3081
% in December 1996 
in a lower state with a flux of $f_{(2 - 10 \rm \, keV)} = 1.3 \times 10^{-12} \rm \, erg \, cm^{-2} \, s^{-1}$ \cite{NGC3081SAX}. Maiolino et al. modeled this spectrum with a Compton thin transmission model with $N_H = 6.4 \times 10^{23} \rm \, cm^{-2}$. Alternatively the data can be modeled by an absorbed ($N_H = 6.6 {+1.8 \atop -1.6} \times 10^{23} \rm \, cm^{-2}$) single power law ($\Gamma = 1.7 {+0.3 \atop -0.4}$) model plus Fe K$\alpha$ line with $EW = 610 {+390 \atop -210} \rm \, keV$ \cite{NGC3081Bassani}.

The measurements of NGC~2992 seem to indicate that the emission of this AGN can be described by the same underlying model throughout the years, but as the main contribution from the continuum can be only measured in the unabsorbed domain at $E \gg 10 \rm \, keV$, additional confirmation in this energy range is required. With the launch of the {\it International Gamma-Ray Astrophysics Laboratory} ({\it INTEGRAL}; Winkler et al. 2003) in October 2002 and {\it Swift} \cite{Swift} in November 2004 we now have access to two hard X-ray telescopes, {\it INTEGRAL} IBIS/SPI and {\it Swift}/BAT, in order to study this behavior. In addition we compare these measurements with observations taken in the last week of the {\it BeppoSAX} mission at the end of April 2002 and in November 1998. As NGC~3081 happens to be in the same field of view for the {\it INTEGRAL} observations, we present the high energy data of both Seyfert galaxies.

The outline of this paper is as follows: In Section~2 we present the data and analysis of the {\it INTEGRAL}, {\it Swift}, and {\it BeppoSAX} data. In Section~3 we discuss the implications of our measurements for the hard X-ray component of NGC~2992 and NGC~3081, and present our conclusions in Section~4.

\section{Data analysis}

\subsection{INTEGRAL data}

{\it INTEGRAL} has observed NGC~2992 for 402.1 ks between 2005-05-04T13:45 and  2005-05-10T06:51 (UTC). The exposure time is the ISGRI effective on-source time. This value is approximately the same for the spectrograph SPI, but the JEM-X and OMC monitors 
cover a much smaller sky area. Thus in the case of dithering observation, as performed for NGC~2992, the source is not always in the field of view of the monitors. The analysis was performed using version 5.1 of the Offline Science Analysis (OSA) software distributed by the {\it INTEGRAL} Science Data Centre (ISDC; Courvoisier et al. 2003). Analysing the {\it INTEGRAL} data altogether, shows that a single power law ($\Gamma = 1.96 {+0.26 \atop -0.23}$) is sufficient to fit the high energy data with $E>20 \rm \, keV$ with a flux of $f_{(20-100 \rm \, keV)} = (6.7 \pm 0.4) \times 10^{-11} \rm \, erg \, cm^{-2} \, s^{-1}$ ($L_X = 8.8 \times 10^{42} \rm \, erg \, s^{-1}$). 
The data of the optical monitor (OMC) show an average brightness of $V = 12.45 \rm \, mag$.   

In the field of NGC~2992 only two additional sources are detected by {\it INTEGRAL} in the X-rays. One is the Seyfert 2 galaxy \object{NGC 3081}. For this source ISGRI data result in a photon index of $\Gamma = 1.8 {+0.3 \atop -0.3}$ with a flux of $f_{(20-100 \rm \, keV)} = (5.8 \pm 0.9) \times 10^{-11} \rm \, erg \, cm^{-2} \, s^{-1}$ ($L_X = 8.1 \times 10^{42} \rm \, erg \, s^{-1}$). This source is not detectable in the SPI data, nor in JEM-X, and no data for NGC~3081 were taken by the OMC.
The second source is the Seyfert 1.9 galaxy \object{MCG -05-23-016}, but its location during the observation was within the partially coded field of view of IBIS, and therefore no reliable information can be extracted for it.

\subsection{Swift data}

{\it Swift} has performed six short pointed observations (duration 1.3 ks  to 11.0 ks) of NGC~2992 between June 14 and July 6, 2006. Because the single XRT spectra showed the same shape and normalisation, we stacked the spectra together, resulting in a total exposure time of 21.5 ks. The {\it Swift}/XRT spectrum is well represented by an absorbed single power law with $\Gamma = 1.01 {+0.15 \atop -0.14}$ and $N_H = 1.59 {+0.97 \atop -0.91} \times 10^{21} \rm \, cm^{-2}$. Note that the absorption includes the Galactic hydrogen column density of $N_{H,gal} = 8 \times 10^{20} \rm \, cm^{-2}$. In addition the iron K$\alpha$ line is detectable with an equivalent width of $EW = 500 \rm \, eV$. Note that this spectrum in the 2--10 keV band represents the sum of different components as discovered by {\it Chandra} \cite{colbert} and is consistent with the same simple power law fit to the {\it Chandra} data.

Apart from the pointed observations, {\it Swift} data are available from the first 15 months of the all-sky survey of the Burst Alert Telescope (BAT, Barthelmy et al. 2005). The extracted spectrum in 4 energy bins ranging from 14 keV to 195 keV can be fit by a single power law model with photon index $\Gamma = 2.04 {+0.34 \atop -0.30}$ and a flux of $f_{(20-100 \rm \, keV)} = (4.5 \pm 0.4) \times 10^{-11} \rm \, erg \, cm^{-2} \, s^{-1}$ ($L_X = 6.0 \times 10^{42} \rm \, erg \, s^{-1}$).

\begin{figure}
%   \centering
   \includegraphics[width=9cm]{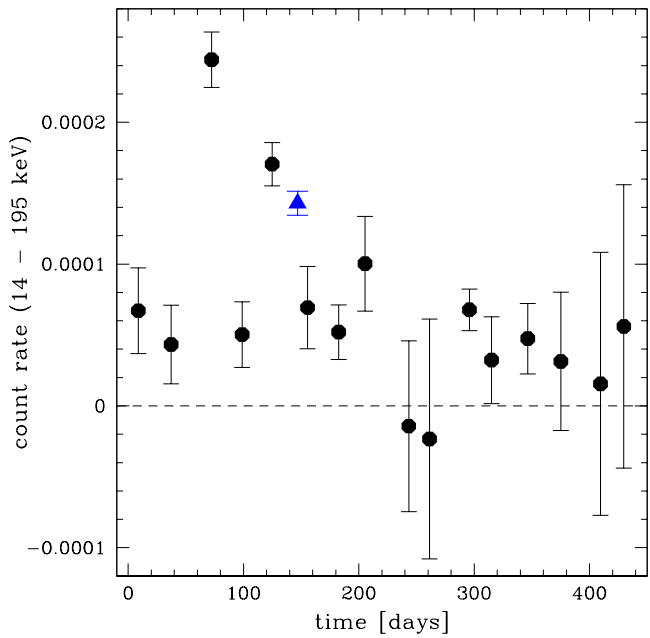}
   \caption{{\it Swift}/BAT 15 month lightcurve of NGC~2992 in on-axis corrected count rate. The triangle indicates the {\it INTEGRAL} IBIS/ISGRI measurement, converted into equivalent BAT count rate. The time indicates the days after the start of the survey in December 11, 2004 at MJD = 53350.15}
              \label{fig:lightcurve}
    \end{figure}

The BAT data can also be used to derive a long-term hard X-ray lightcurve. In order to do so, the data were binned in time in 28 day intervals, as shown in Fig.~\ref{fig:lightcurve}. The lightcurve is inconsistent with a constant flux on a $>10 \sigma$ level, indicating that indeed NGC~2992 exhibits strong variability, up to a factor of 6 at time scales as short as a month. The count rates have been corrected for off-axis measurements and for partially coding and thus reflect the count rate the source would have in an on-axis observation. Based on the BAT data we derived hardness ratios $HR = (H-S)/(H+S)$ with count rates in the bands $S$ (14 - 24 keV) and $H$ (24 - 195 keV).  For the period in which the BAT count rate was in a high state with a total count rate of $CR_{14-195 \rm \, keV} > 10^{-4} \rm \, counts \, s^{-1}$, the hardness ratio is $HR = 0.33 \pm 0.06$, and in the low state $HR = 0.24 \pm 0.07$. We therefore do not detect significant variability of the spectral shape in the {\it Swift}/BAT data.
%As no systematic error has been added here, we do not consider this difference as statistically significant. 
This is also shown when analysing the 4 channel spectra in high and low state separately, which results in the same spectral slope. 

\begin{figure}
%   \centering
   \includegraphics[width=9cm]{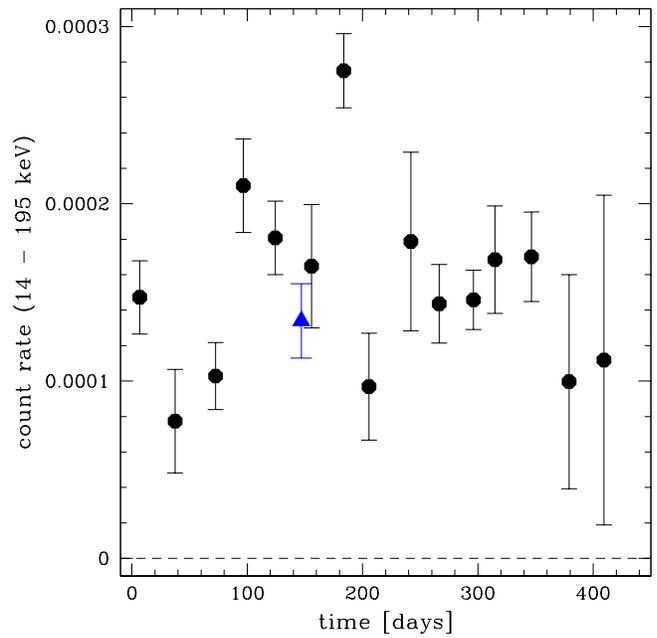}
   \caption{{\it Swift}/BAT 15 month lightcurve of NGC~3081 in on-axis corrected count rate. The triangle indicates the {\it INTEGRAL} IBIS/ISGRI measurement, converted into equivalent BAT count rate. The time indicates the days after the start of the survey on December 11, 2004 at MJD = 53350.15}
              \label{fig:lightcurveNGC3081}
    \end{figure}

The BAT spectrum of NGC~3081 can also be modeled by a simple power law with photon index $\Gamma = 1.72 {+0.23 \atop -0.22}$ and a flux of $f_{(20-100 \rm \, keV)} = (6.9 \pm 0.5) \times 10^{-11} \rm \, erg \, cm^{-2} \, s^{-1}$ ($L_X = 9.7 \times 10^{42} \rm \, erg \, s^{-1}$). The light curve extracted from the BAT survey  is inconsistent with a constant flux on a $>6 \sigma$ level, showing variability by a factor of 7 (Fig.~\ref{fig:lightcurveNGC3081}). The average hardness ratio for the high state ($CR_{14-195 \rm \, keV} > 1.5 \times 10^{-4} \rm \, counts \, s^{-1}$) is $HR = 0.43 \pm 0.06$ and  $HR = 0.26 \pm 0.07$ in the low state.


\subsection{BeppoSAX data}

One of the last observations performed by {\it BeppoSAX} was done on April 26, 2002 targeting NGC~2992. At that time only the PDS instrument (15 - 200 keV) was operational. The 18.5 ks of data show a simple power law spectrum with photon index $\Gamma = 1.92 {+0.08 \atop -0.08}$ and a flux of $f_{(20-100 \rm \, keV)} = (1.53 \pm 0.03) \times 10^{-10} \rm \, erg \, cm^{-2} \, s^{-1}$. 
%In order to make use of all available high energy data, 
We also analysed the other available {\it BeppoSAX}/PDS observations of NGC~2992. One had been performed on November 25, 1998 with a total effective exposure time of 27.1 ks. The spectral fit shows a similar spectrum with $\Gamma = 1.94 {+0.10 \atop -0.10}$ and  $f_{(20-100 \rm \, keV)} = (1.16 \pm 0.03) \times 10^{-10} \rm \, erg \, cm^{-2} \, s^{-1}$. Another 33.5 ks {\it BeppoSAX} observation took place December 1st, 1997, and the spectrum shows $\Gamma = 2.2 {+1.2 \atop -0.7}$ and $f_{(20-100 \rm \, keV)} = (1.4 \pm 0.3) \times 10^{-11} \rm \, erg \, cm^{-2} \, s^{-1}$. {\it BeppoSAX} observed NGC~3081 on December 20, 1996, but only 2 ks of PDS data were accumulated which did not allow a detection..

\begin{table*}
\caption{The 20 - 100 keV spectrum of NGC~2992}             % title of Table
\label{measurements}      % is used to refer this table in the text
\centering                          % used for centering table
\begin{tabular}{c c c c c}        % centered columns (4 columns)
\hline\hline                 % inserts double horizontal lines
Instrument & Start Observation  & Exposure time & flux & $\Gamma$ \\    % table heading 
           &  & [ks] & $[10^{-11} \rm \, erg \, cm^{-2} \, s^{-1}]$ & \\
\hline                        % inserts single horizontal line
  {\it BeppoSAX}/PDS & 01.12.1997 & 33.5 &  $1.4 \pm 0.3$ &  $2.2 {+1.2 \atop -0.7}$ \\  
  {\it BeppoSAX}/PDS & 25.11.1998 & 27.1 & $11.6 \pm 0.3$ &  $1.94 {+0.10 \atop -0.10}$ \\
  {\it BeppoSAX}/PDS & 26.04.2002 & 18.5 & $15.3 \pm 0.3$ & $1.92 {+0.08 \atop -0.08}$\\
  {\it Swift}/BAT  & 11.12.2004 & 9 month survey & $4.5 \pm 0.4$ & $2.04 {+0.34 \atop -0.30}$\\
  {\it INTEGRAL}/IBIS  & 04.05.2005 & 402.1 & $6.7 \pm 0.4$ & $1.96 {+0.26 \atop -0.23}$\\
\hline                                   %inserts single line
\end{tabular}
\end{table*}

\subsection{Combined data analysis}

\begin{figure}
%   \centering
   \includegraphics[width=9cm]{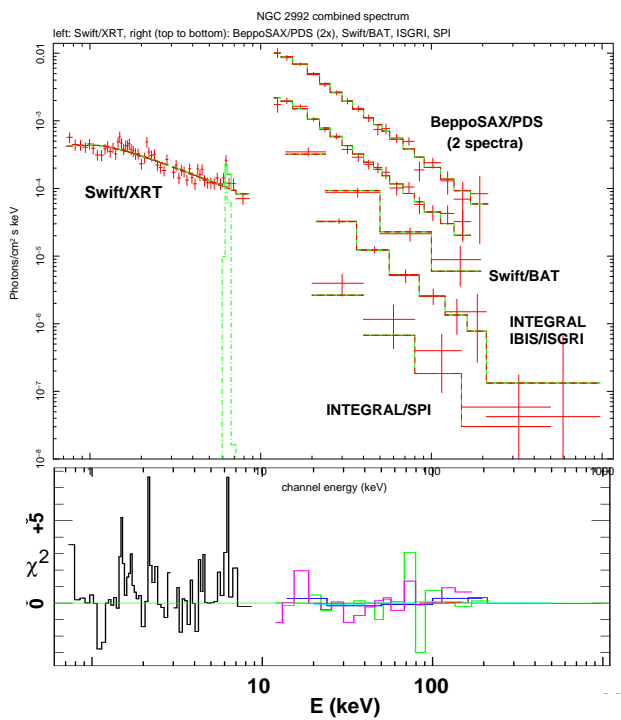}
   \caption{Combined spectrum of NGC~2992 based on {\it BeppoSAX},
     {\it Swift} and {\it INTEGRAL} data in photon space. The same
     model has been applied to all data, plus intercalibration factors
     in order to account for flux variations. Note that except for the
     {\it INTEGRAL} IBIS/ISGRI and the {\it Swift}/XRT spectrum, all
     the spectra have been offset in this plot from their original
     position for better readability. The {\it BeppoSAX} spectrum from
     1997 is not shown. The lower panel shows the contribution in each
     energy bin to the $\chi^2$ value, multiplied by the sign of the residual.}
              \label{FigGam}
    \end{figure}

Comparing the different data sets at $E > 20 \rm \, keV$ taken of
NGC~2992 over the years by three different missions and four different
instruments, it is apparent that although the flux level of the
continuum varied, the spectral slope stayed constant within the errors
of the individual measurements. A list summarizing the individual
measurements at $E > 20 \rm \, keV$ is given in
Table~\ref{measurements}. We therefore performed a fit of all data
simultaneously, applying intercalibration factors in order to account
for different flux levels but using the same model. The resulting
spectrum is shown in Figure~\ref{FigGam}. The overall fit shows an
absorbed broken power law. The modeled absorption of $N_H = 1.58
{+0.90 \atop -0.86} \times 10^{21} \rm \, cm^{-2}$ is mainly dominated
by the XRT spectrum, as is the spectral slope of $\Gamma_1 = 0.99
{+0.14 \atop -0.06}$ below the turnover of $E_{break} \simeq 16 \rm \, keV$.
%$E_{break} = 15.8 {+1.9 \atop -3.5} \rm \, keV$. 
It has to be considered that the exact
location of the break energy cannot be given, as no strictly
simultaneous data are used between the {\it Swift}/XRT and the high
energy data. Nevertheless a break between 10 keV and 20 keV in the
spectrum is necessary. Above this energy the spectrum is described by a power law of $\Gamma_2 = 1.97 {+0.07 \atop -0.07}$. The fit results in a $\chi_\nu^2 = 0.98$ for 103 degrees of freedom, while a simple power law fit gives an unacceptable $\chi_\nu^2 = 2.1$ (107 d.o.f.). 
The addition of a high-energy cut off in the spectrum does not improve the fit. %On the contrary, a cut-off at an energy below $E_C = 250 \rm \, keV$ can be rejected on a $1\sigma$ level.

\begin{figure}
%   \centering
   \includegraphics[width=9cm]{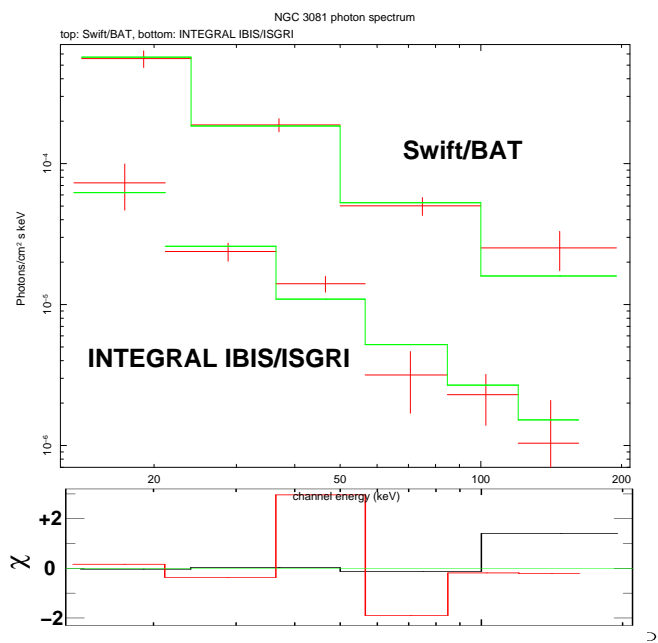}
   \caption{{\it Swift}/BAT and {\it INTEGRAL} IBIS/ISGRI spectrum of
     the Seyfert 2 galaxy NGC~3081. The BAT spectrum has been
     multiplied by a factor of 10 for better readability. The single
     power law model fit results in a photon index of $\Gamma = 1.8
     {+0.2 \atop -0.2}$. The lower panel shows the contribution in each
     energy bin to the $\chi^2$, multiplied by the sign of the residual..}
              \label{fig:NGC3081}
    \end{figure}

The {\it INTEGRAL} IBIS/ISGRI and {\it Swift}/BAT data also show the same spectral parameters. A combined fit to both data sets, with variable normalisation, results in a photon index of $\Gamma = 1.76 {+0.19 \atop -0.18}$, where the BAT data show a 6\% larger flux compared to the IBIS/ISGRI data. The combined plot is shown in Figure~\ref{fig:NGC3081}. Addition of a high-energy cut off does not improve the fit but the data do not allow to constrain a possible cut-off.


\section{Discussion}

The hard X-ray spectrum of NGC~2992 can be described by the same simple power law model with photon index $\Gamma \simeq 2$ throughout all the different observations performed over the last thirty years. All measurements show the same power law index of $\Gamma \simeq 2$ while the intensity varies by a factor of 11, according to the long-term variability, and by a factor of 6, using the one-month time binned lightcurve of the {\it Swift}/BAT survey. The flat slope of $\Gamma \simeq 1$ as observed by {\it Swift}/XRT in the 2--10~keV range has been shown to be the superposition of three individual components (Colbert et al. 2005, Yaqoob et al. 2007). The hard X-ray spectrum can also be studied by {\it Suzaku}'s collimated Hard X-ray Detector (HXD), but as the background model for this instrument is still under development, no conclusive results can be drawn for NGC~2992 from {\it Suzaku} data yet (Yaqoob et al. 2007). 
A monitoring campaign in the 2~--~10~keV energy range performed by the PCA on board the {\it Rossi X-ray Timing Explorer} ({\it RXTE}) between March 2005 and January 2006 reveiled the same behavior as seen in the $> 20 \rm \, keV$ data, i.e. that the intensity of the underlying continuum varies, whereas the spectral slope stays constant within the statistical error (Murphy, Yaqoob, \& Terashima 2007). Also in this energy band the luminosity varied by a factor of 11, whereas the slope of the power-law fit was consistent with no variability (weighted mean $\Gamma \simeq 1.7$). The timescale of those variations is of the order of days to weeks, which leads Murphy et al. to the suggestion that this variability is based on short-term flaring activity as opposed to long-term changes in accretion activity.

A similar behavior, of a stable photon index of the unabsorbed hard X-ray spectrum under varying intensity, has been observed already in two other well-studied objects, \object{NGC 4388} \cite{NGC4388} and \object{NGC 4151} \cite{NGC4151}.
What kind of mechanism can produce such a constant shape of the hard
X-ray emission while the intensity varies significantly? If we assume
that the hard X-ray emission originates from hot plasma (e.g. a hot
extended corona above the relatively cold accretion disk; e.g. Haardt
\& Maraschi 1991), the temperature of this plasma has apparently been
constant. The variations in intensity can then be explained in two
ways: either the amount of material emitting the hard X-rays varies,
or the amount of plasma visible to the observer varied. The latter can be caused either through absorption near the central engine of the AGN, or through different orientation of the disk with respect to the observer, or a mixture of both effects. The intrinsic absorption measured in soft X-ray does not exhibit large variations ($N_H = (0.2 - 1.0) \times 10^{22} \rm \, cm^{-2}$) and this absorption has no effect on the hard X-ray spectrum. In addition a variable Compton-thick absorber, which would be able to change the flux at $> 20 \rm \, keV$ significantly, would also effect the spectral slope, which is not observed. Different orientation of the accretion disk with respect to the observer also should not have an influence on the hard X-ray emission, as the disk is thought to be Compton-thin.

This leaves us with the possibility of variable amount of hard X-ray
emitting material. An explanation for this 
is provided by a model including hot flares produced via magnetic field reconnections (e.g. Galeev et al. 1979; Poutanen \& Fabian 1999). In this flare model, sudden dissipation occurs in very localized regions above the disk in coronal loops. It has to be pointed out though, that the expected duration of those flares is rather on the scale of a day than on a monthly basis (Czerny et al. 2004). A short flare would most likely not be detectable in the {\it Swift}/BAT survey.  

NGC~2992 and NGC~3081 are two more examples for AGN with hard X-ray
spectra which do not require to model a cut-off in the spectrum. Out of 36 Seyfert type AGN in the first {\it INTEGRAL} catalog \cite{intagn} only six objects required a high energy cut-off in the spectrum in the range 35 keV to 181 keV in order to achieve a reasonable spectral fit. 
The two objects presented here give further evidence that the hard X-ray spectrum of Seyfert galaxies extends as a power-law up to hardest X-ray energies. A possible cut-off in the spectra appears at energies $E_C \gg 100 \rm \, keV$, outside the range where  {\it INTEGRAL} and {\it Swift} provide sufficient sensitivity for most of the Seyfert galaxies.

\section{Conclusions}

Two Seyfert galaxies are detectable in a 400 ks observation by {\it INTEGRAL} IBIS/ISGRI pointed at NGC~2992. The target itself and the Seyfert 2 galaxy NGC~3081. Both sources are also detected by {\it Swift}/BAT. The results can be summarized as follows:

   \begin{enumerate}
      \item NGC~2992: This well-studied Seyfert 1.9 galaxy exhibits a hard X-ray spectrum with constant photon index $\Gamma \simeq 2$ and variable $20 - 100 \rm \, keV$ luminosity, ranging from $L_X = 1.2 \times 10^{43} \rm \, erg \, s^{-1}$ to $L_X = 3.9 \times 10^{43} \rm \, erg \, s^{-1}$. The {\it Swift}/BAT survey shows that the flux of NGC~2992 is variable by a factor of 6 on time scales of a month.
%at least, where the variability can be as large as a factor of 6. 
The source exhibits a constant spectral shape ($\Gamma \simeq 2$) over a period of 7 years, as shown by {\it BeppoSAX}/PDS, {\it INTEGRAL}/IBIS, and {\it Swift}/BAT, while varying in intensity by an order of magnitude.
This result is consistent with variability studies in the $2 - 10 \rm \, keV$ range, which also detected strong flux variations by a factor of 11 with a constant spectral slope (Murphy, Yaqoob, \& Terashima 2007).% different amount of hot plasma visible to the observer (caused by either variable absorption or geometrical effects such as a precession of the accretion disk), rather than by temperature changes. 
      \item NGC~3081: For this Seyfert 2 galaxy we presented for the first time a hard X-ray spectrum above 20 keV based on {\it INTEGRAL} IBIS/ISGRI and {\it Swift}/BAT measurements. The spectrum can be modeled by a single power law with $\Gamma = 1.8 {+0.2 \atop -0.2}$ and the source shows a luminosity of $L_X = 8.1 \times 10^{42} \rm \, erg \, s^{-1}$. The {\it Swift}/BAT survey shows significant variability up to a factor of 7, without significant variation of the spectral slope.
   \end{enumerate}

The {\it INTEGRAL}/IBIS and {\it Swift}/BAT data of NGC~2992 show
strong evidence for a non-variable spectral slope on both, short and
long time scales. Also NGC~3081 does not show spectral variability on
monthly time scales in the {\it Swift}/BAT data. In both cases the
luminosity varies by an order of magnitude. Even though NGC~3081 shows
significant absorption, it is Compton thin
\citep{NGC3081Bassani,NGC3081SAX}, thus in both Seyfert 2 galaxies
variable absorption cannot account for strong flux variability at
energies $>20 \rm \, keV$. Even strong flux variations on a monthly
time scale do not show changes in the spectral slope. This is further
evidence that absorption does not play a role here, as column
densities necessary to decrease the flux by an order of magnitude
(i.e. $N_H \gg 10^{24} \rm \, cm^{-2}$) would flatten the spectrum significantly. 
We therefore suggest that hot flares are responsible for the flux
changes, as described e.g. in Galeev et al. (1979), Poutanen \& Fabian
(1999), and Czerny et al. (2004). In this model long lasting flares occur in
the accretion disk which increase the amount of hot plasma, while the
temperature of the X-ray emitting matter stays constant. 

Further investigations of the hard X-ray spectrum in more AGN is
required in order to see whether all Seyfert galaxies show a constant
spectral shape, or if NGC~2992, NGC~4388, and NGC~4151 are special
cases. Especially the {\it Swift}/BAT all-sky survey will provide an
excellent data base for this kind of studies. A preliminary analysis
of the variability properties of the 45 brightest AGN seen in {\it
  Swift}/BAT (NGC~3081 is not among those sources) shows that NGC~2992 exhibits the strongest flux
variability of the Seyfert galaxies (Beckmann et al. in prep.).

%\begin{acknowledgements}
%This research has made use
% of the Tartarus (Version 3.1) database, created by Paul O'Neill and Kirpal Nandra at Imperial College London, and Jane Turner at NASA/GSFC. Tartarus is supported by funding from PPARC, and NASA grants NAG5-7385 and NAG5-7067.
%\end{acknowledgements}

\end{document}